%% file: csbec.tex
\documentclass[pra,twocolumn,showpacs]{revtex4}

\usepackage{epsf}

\newcommand{\x}{\tilde{x}}
\newcommand{\z}{\tilde{t}}
\newcommand{\ii}{\mathrm{i}}

\begin{document}

\title{Complex scaling approach to the decay of Bose-Einstein condensates}

\author{Peter Schlagheck}
\author{Tobias Paul}
\affiliation{Institut f\"ur Theoretische Physik, Universit\"at Regensburg,
  93040 Regensburg, Germany}

\date{\today}

\begin{abstract}

The mean-field dynamics of a Bose-Einstein condensate is studied in presence 
of a microscopic trapping potential from which the condensate can escape via
tunneling through finite barriers.
We show that the method of complex scaling can be used to obtain a
quantitative description of this decay process.
A real-time propagation approach that is applied to the complex-scaled
Gross-Pitaevskii equation allows us to calculate the chemical potentials and
lifetimes of the metastably trapped Bose-Einstein condensate.
The method is applied to a one-dimensional harmonic confinement potential
combined with a Gaussian envelope, for which we compute the lowest symmetric
and antisymmetric quasibound states of the condensate.
A comparison with alternative approaches using absorbing boundary conditions
as well as complex absorbing potentials shows good agreement.

\end{abstract}


\maketitle

\section{Introduction}

\input{csbec_in.tex}

\section{Complex scaling of the Gross-Pitaevskii equation}

\input{csbec_gp.tex}

\section{Numerical results}

\input{csbec_rs.tex}

\section{Conclusion}

\input{csbec_cc.tex}

\begin{acknowledgments}

We are indebted to Hans-J\"urgen Korsch for fruitful and inspiring discussions
which initiated the project. Valuable suggestions from Klaus Richter,
Nimrod Moiseyev, Joachim Brand, Sandro Wimberger, and Tsampikos Kottos are
gratefully acknowledged.

\end{acknowledgments}

\input{csbec.bbl}
\end{document}

%% file: csbec_in.tex
With the advent of optical lattices and ``atom chips'' \cite{FolO00PRL}, it
became possible to probe the transport properties of a Bose-Einstein
condensate in the mesoscopic regime.
Experiments in this context include the observation of Bloch oscillations
\cite{AndKas98S,MorO01PRL}, the guided and free propagation of condensates
through waveguide structures \cite{OttO01PRL,HaeO01PRL}, the transport of
condensates with ``optical tweezers'' \cite{GusO02PRL}, as well as the
realization of Josephson junctions \cite{AlbO05PRL} and matter-wave
interferometry \cite{SchO05N}, to mention just a few examples.
Those experiments typically involve rather small trapping potentials, with
length scales that can be of the order of a few microns.
In such microscopic geometries, \emph{decay} mechanisms of the condensate
become a relevant issue.
On the one hand, the condensed state is, at finite atom densities, subject to
depletion, which is caused by the interaction with the thermal cloud and by
three-body collisions.
On the other hand, the condensate can escape from the trapping potential by
tunneling though its barriers, if the chemical potential of the condensed 
atoms exceeds the background potential in the free space outside the trap.
In that case, the self-consistent mean-field state of the condensate is no
longer bound, but rather corresponds to a metastable ``resonance'' state, in a
similar way as, e.g., doubly excited electronic states in the helium atom.

From the theoretical point of view, the problem of metastable states of
Bose-Einstein condensates in such ``open'' trapping potentials was first
approached by Moiseyev et al.\ in Ref.~\cite{MoiO03JPB}.
In this work, the mean-field dynamics of a condensate was studied in presence
of an isotropic harmonic confinement that is combined with a Gaussian
envelope.
Decaying states of the condensates were identified with self-consistent
solutions of the stationary Gross-Pitaevskii equation at complex-valued
chemical potentials, where complex absorbing potentials at the boundaries of
the numerical grid were used to account for the escape of the condensate.
As a result, a cross-over from a decaying (resonance) state to a bound state
of the condensate was found for a finite attractive interaction between the
atoms. 

The calculations of Ref.~\cite{MoiO03JPB} were satisfactory from the
quantitative point of view, but raised an important open question:
How can ``resonances'', i.e.\ stationary states that describe the escape of
population from an open confinement potential, be formally introduced in the
framework of the nonlinear Gross-Pitaevskii equation?
For linear systems, it is well known that this task is most conveniently
accomplished by applying the method of ``complex scaling'' (or ``complex
rotation'') \cite{BalCom71CMP,Sim73AM,Rei82ARPC,Moi98PR}.
This technique essentially amounts to the complex dilations
$\mathbf{r} \mapsto \mathbf{r} e^{i \theta}$ and $- \ii \nabla \mapsto - \ii \nabla e^{-i \theta}$ of the
position and momentum operators in the Hamiltonian that describes the quantum
system under study.
This transformation leads to a nonhermitean Hamiltonian with a complex
eigenvalue spectrum the continuous part of which is rotated to the lower half
of the complex energy plane.
Resonances, i.e.\ decaying states with eigenvalues corresponding to poles of
the resolvent below the real energy axis, are thereby uncovered and can
be calculated using standard diagonalization techniques for complex matrices.
This approach is essentially \emph{exact}, in the sense that no a priori
approximations are introduced in the complex dilation procedure.
Positions and widths of resonances can therefore be calculated with high
precision by means of the complex scaling procedure \cite{Rei82ARPC,Moi98PR}.

Quite obviously, the extension of the method to the Gross-Pitaevskii equation
is far from straightforward (see, e.g., the discussion in
Ref.~\cite{CedTarWin90JPB}).
Indeed, the presence of the nonlinear term inhibits a direct formulation of
complex scaling in terms of Green's functions or resolvents, which, for linear
systems, provide the convenient link to the time-dependent decay problem.
Furthermore, self-consistent solutions of the Gross-Pitaevskii equation are
usually defined with respect to a given \emph{normalization} of the
condensate wavefunction $\psi$ --- which poses a conceptual problem for decaying
states that are generally non-normalizable.
An additional complication (which is quite severe from the numerical point of
view, as we shall see later on) arises from the fact that the nonlinear term
in the Gross-Pitaevskii contains the square modulus of $\psi$ and is therefore a
\emph{nonanalytic} function of the condensate wavefunction.

This last difficulty was circumvented in a recently published approach by
Moiseyev and Cederbaum \cite{MoiCed05PRA}.
In this work, the complex scaling transformation was applied to the exact
microscopic many-particle dynamics of the bosonic system, and a variational
ansatz was used to derive from there a complex-scaled version of the
Gross-Pitaevskii equation.
This variational ansatz, however, was implicitly based on the assumption that
the condensate wavefunction of the decaying state is entirely \emph{real},
which effectively means that $|\psi(\mathbf{r})|^2$ can be replaced by the
analytic term $\psi^2(\mathbf{r})$ in the Gross-Pitaevskii equation.
Moiseyev and Cederbaum showed that chemical potentials and decay rates can be
calculated from this complex nonlinear Schr\"odinger equation, whose behaviour
as a function of the interaction strength seems to be in qualitative agreement
with the intuitive expectation.

Our ansatz in this paper is substantially different in the sense that we
explicitly take into account the possibility of \emph{complex-valued}
resonance wavefunctions.
As pointed out above, this introduces a major complication of the problem,
due to the presence of the $|\psi(\mathbf{r})|^2$ term in the Gross-Pitaevskii
equation.
We solve this problem by formally defining separate dilations of $\psi$ and its
complex conjugate, and by performing explicit complex rotations of the
wavefunction in the numerical implementation.
Our approach is supported by calculations based on the ``real'', i.e.\
unscaled, Gross-Pitaevskii equation in presence of absorbing boundaries, with
which we obtain good agreement. 

The paper is organized as follows:
In Section \ref{s:p}, we derive the relation between complex ``resonance''
solutions of the stationary Gross-Pitaevskii equation and the actual
time-dependent decay process of the condensate.
For the sake of simplicity, we restrict ourselves to the case of the
one-dimensional mean-field dynamics of a Bose-Einstein condensate in a
magnetic or optical waveguide. 
The complex scaling method and its application to the Gross-Pitaevskii
equation is explained in Section \ref{s:c}, and details on the numerical
implementation of the method are presented in Section \ref{s:n}.
Section \ref{s:r} contains the discussion of the numerical results that we
obtain for a harmonic trapping potential with Gaussian envelopes.

%% file: csbec_gp.tex
\subsection{The time-dependent decay problem}

\label{s:p}

\begin{figure}
\begin{center}
\leavevmode
\epsfxsize8.6cm
\epsfbox{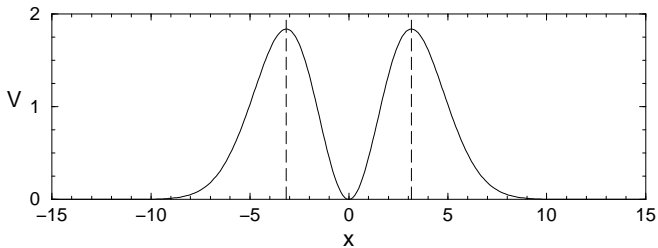}
\end{center}
\caption{Effective longitudinal potential (\ref{eq:v}) for $\alpha = 0.1$. 
  The vertical dashed lines define the spatial region within which the
  wavefunction is normalized according to Eq.~(\ref{eq:norm}).
\label{fg:pot}
}
\end{figure}

We start from the one-dimensional time-dependent Gross-Pitaevskii equation
\begin{eqnarray}
 \ii \hbar \frac{\partial}{\partial \z} \Psi(\x,\z) & = & 
 - \frac{\hbar^2}{2m} \frac{\partial^2}{\partial \x^2} \Psi(\x,\z)
 + V(\x) \Psi(\x,t) \nonumber \\
 & & + 2 a_s \hbar \omega_\perp |\Psi(\x,\z)|^2 \Psi(\x,\z) \label{eq:GP} 
\end{eqnarray}
which describes the mean-field dynamics of the condensate in presence of a
tight cylindrical confinement with transverse frequency $\omega_\perp$. 
Here, $m$ is the mass and $a_s$ the $s$-wave scattering length of the
atoms.
For the sake of definiteness, we consider, as in Ref.~\cite{MoiO03JPB}, the
open longitudinal potential
\begin{equation}
  V(\x) = \frac{1}{2} m \omega_0^2 \x^2 \exp( - \tilde{\alpha} \x^2 ) \, , \label{eq:V}
\end{equation}
which is shown in Fig.~\ref{fg:pot}.
Variants of (\ref{eq:V}) were extensively studied in the literature on
resonances (e.g.\ \cite{MoiCerWei78MP,RitElaBra81PRA,KorGlu02EJP}).

Since $V(\x) \to 0$ for $\x \to \pm \infty$, the system does not exhibit any bound state.
For not too large values of $\tilde{\alpha}$, however, the condensate can be
temporarily stored within the potential well, from where it decays via
tunneling through the barriers.
In the linear case of noninteracting atoms ($a_s = 0$), this decay process is
described by resonance states $\Psi(\x,\z) = \Psi(\x) \exp( - \ii E \z / \hbar)$ where
$\Psi(\x)$ satisifies the stationary Schr\"odinger equation for the complex
energy $E = \mu - \ii \Gamma/2$ and exhibits outgoing (Siegert) boundary conditions
\cite{Sie39PR} of the form $\Psi(\x) \longrightarrow \Psi_0 \exp(\ii k |x|)$ with ${\rm Re}(k) >
0$ for $x \to \pm \infty$ .
An exponential decay $\propto \exp( - \Gamma \z / \hbar)$ is therefore obtained for the atomic
density inside the well if the system is initially prepared in such a
resonance state.
This is fundamentally different in presence of finite interaction ($a_s \neq 0$)
where the tunnel coupling through the barriers explicitly depends, via the
nonlinear term in the Gross-Pitaevskii equation, on the density
$|\Psi(\x,\z)|^2$.
As a consequence, the decay rate $\Gamma$ varies during the time evolution of the
condensate, which results in a nonexponential decay (see, e.g.,
\cite{WimO05xxx}).

As long as the rate $\Gamma$ characterizing the temporal variation of the density
inside the well is comparatively small, the decay process can be approximately
described by an adiabatic ansatz where the condensate is assumed to remain 
always in the energetically lowest (and most stable) resonance state
associated with a given instantaneous density $|\Psi(\x,\z)|^2$.
To this end, we introduce dimensionless variables which are formally obtained
by setting $\hbar = m = \omega_0 = 1$.
Specifically, we define the dimensionless potential
\begin{equation}
  v(x) = V(a_0 x) / \hbar \omega_0 = \frac{1}{2} x^2 \exp( - \alpha x^2 ) \label{eq:v}
\end{equation}
with the dimensionless position $x = \x / a_0$ and $\alpha = a_0^2 \tilde{\alpha}$,
where $a_0 = \sqrt{\hbar / (m \omega_0)}$ is the oscillator length associated with the
well (which would be of the order of several microns for typical experimental
setups).
We furthermore define, as a function of the effective interaction strength
$g$, the dimensionless resonance state $\psi_g(x)$, together with its associated
eigenvalue $E_g = \mu_g - \ii \Gamma_g / 2$, as the solution of the stationary equation
\begin{equation}
  H(\psi_g) \psi_g(x) = E_g \psi_g(x) \label{eq:gp}
\end{equation}
with the dimensionless nonlinear Hamiltonian
\begin{equation}
  H(\psi) = -\frac{1}{2} \frac{\partial^2}{\partial x^2} + v(x) + g |\psi(x)|^2 \, .
  \label{eq:h}
\end{equation}
In addition to Eq.~(\ref{eq:gp}), we demand that $\psi_g$ be normalized according
to the condition
\begin{equation}
  {\mathcal N}[\psi_g] \equiv \int_{-\infty}^\infty |\psi_g(x)|^2 w(x) dx = 1 \label{eq:norm0}
\end{equation}
where the weight function $w(x)$ measures the population inside the well.
We shall use $w(x) = \theta( x_\alpha - x ) \theta( x + x_\alpha )$ in the following, i.e.,
\begin{equation}
  {\mathcal N}[\psi_g] = \int_{-x_\alpha}^{x_\alpha} |\psi_g(x)|^2 dx \, , \label{eq:norm}
\end{equation}
where $x_\alpha = 1 / \sqrt{\alpha}$ corresponds to the maximum of the barrier.
We remark that the numerical results presented in Section \ref{s:r} do not
sensitively depend on the particular choice of $x_\alpha$.

A further requirement that one should impose are outgoing boundary conditions
for the wavefunction.
Due to the selfconsistent nonlinear term in the Gross-Pitaevskii equation,
however, this requirement cannot be formulated in the same explicit way as for
the linear case \cite{Sie39PR}.
Qualitatively, such outgoing boundary conditions imply that the local current
density of $\psi_g(x)$ is directed away from the trapping potential, and that no
additional back-reflections arise outside the support of $v(x)$.
This means that the wavefunction evolves according to
$\psi_g(x) \propto \exp[ \ii \int^x k(x') dx']$ where the real part of the effective wave
number $k(x)$ is positive (negative) for $x \gg x_\alpha$ ($x \ll -x_\alpha$) and varies
smoothly with $x$ to account for the variation of the selfconsistent potential
in Eq.~(\ref{eq:h}).
We shall assume that such a condition is met for the wavefunction $\psi_g$.

The adiabatic ansatz for $\Psi(\x,\z)$ is now formulated as
\begin{equation}
  \Psi(\x,\z) = \sqrt{\frac{N_0}{a_0}} \psi_{g(t)}(x)
  \exp\left( - \ii \int_0^{t} E_{g(t')} dt' \right) \label{eq:psi_ad}
\end{equation}
with $x \equiv \x / a_0$ and $t \equiv \omega_0 \z$, where the effective time-dependent
interaction strength is given by
\begin{equation}
  g(t) = 2 \frac{a_s}{a_0}\frac{\omega_\perp}{\omega_0} N(t) \, . \label{eq:gt}
\end{equation}
$N(t)$ is the time-dependent population inside the well and decays according
to the equation
\begin{equation}
  \frac{d N}{d t} = - \Gamma_{g(t)} N(t) \label{eq:dN}
\end{equation}
with the initial condition $N(0) = N_0$, which is formally solved yielding
\begin{equation}
  N(t) = N_0 \exp\left( - \int_0^t \Gamma_{g(t')} d t' \right) \, . \label{eq:N}
\end{equation}
As can be straightforwardly verified, $\Psi(\x,\z)$ satisfies the time-dependent
Gross-Pitaevskii equation (\ref{eq:GP}) as long as the temporal variation of
$\psi_{g(\omega_0t)}(\x/a_0)$ (which is proportional to $\Gamma_g$) can be neglected in
Eq.~(\ref{eq:GP}).
We thereby obtain
\begin{equation}
  \int_{-\infty}^\infty |\Psi(\x,\z)|^2 w(\x/a_0) d\x = N(\omega_0 \z) \, .
\end{equation}

The full characterization of the time-dependent decay process within this
adiabatic picture therefore requires to calculate the resonance states $\psi_g$
and their complex energies $E_g$ within the range $0 < g < g(0)$ for $a_s >
0$, and within $g(0) < g < 0$ for $a_s < 0$.
We shall show now that this aim can be achieved with the method of complex
scaling.
To simplify the discussion, we shall drop the index $g$ (i.e., $\psi \equiv \psi_g$ and 
$E \equiv E_g$) and consider $\alpha = 0.1$ in the following.

\subsection{The complex scaling transformation}

\label{s:c}

\subsubsection{Linear case}

Our aim is now to calculate the resonance states for the nonlinear
Gross-Pitaevskii equation --- i.e, the solutions of
\begin{equation}
  H(\psi) \psi(x) = E \psi(x) \label{eq:gp2} \, ,
\end{equation}
with
\begin{eqnarray}
  H(\psi) & = & H_0 + g |\psi(x)|^2 \, , \label{eq:h2} \\
  H_0 & = & -\frac{1}{2} \frac{\partial^2}{\partial x^2} + v(x) \, ,
\end{eqnarray}
that exhibit outgoing boundary conditions and are normalized according to
\begin{equation}
  \int_{-x_\alpha}^{x_\alpha} |\psi(x)|^2 dx = 1 \, . \label{eq:norm2}
\end{equation}
For the linear case, i.e.\ in the absence of interaction ($g=0$), this aim
can be straightforwardly achieved by the method of complex dilation.
This operation is formally defined by the linear mapping
\begin{equation}
  \psi(x) \mapsto \psi^{(\theta)}(x) \equiv R_\theta \psi (x) = e^{\ii \theta / 2 } \psi( x e^{\ii \theta} ) \label{eq:cs}
\end{equation}
which effectively amounts to a nontrivial scaling of the configuration space
by the complex factor $\exp(\ii \theta)$.
Applying this transformation to the stationary Schr\"odinger equation 
(\ref{eq:gp2}) (for $g=0$) yields
\begin{equation}
  H_0^{(\theta)} \psi^{(\theta)}(x) = E \psi^{(\theta)}(x) \label{eq:sgrot}
\end{equation}
where
\begin{equation}
  H_0^{(\theta)} \equiv R_\theta H_0 R_\theta^{-1} = - \frac{1}{2} e^{-2 \ii \theta} \frac{\partial^2}{\partial x^2} +
  v(x e^{i \theta}) \label{eq:h0rot}
\end{equation}
represents the (nonhermitean) complex scaled Hamiltonian.

The spectral properties of the complex Schr\"odinger equation (\ref{eq:sgrot})
are widely discussed in the literature on complex scaling
\cite{BalCom71CMP,Sim73AM,Rei82ARPC,Moi98PR}.
While all eigenstates of the real equation (\ref{eq:gp2}) will formally solve
also Eq.~(\ref{eq:sgrot}) after the transformation, their normalizability
properties might change under the operation (\ref{eq:cs}):
Bound states of $v(x)$ (which are absent in our particular case) fall off
sufficiently rapidly with increasing $|x|$, such that they remain normalizable
under complex dilation (as long as $|\theta| < \pi/4$).
This is, however, not true for continuum states, which turn into wavefunctions
that exponentially diverge for $|x| \to \infty$.
For compensation, a ``new'' continuum of asymptotically oscillatory states
arises in the rotated system, with eigenvalues along the axis 
$E = \epsilon e^{-2 \ii \theta}$ with real $\epsilon$.

\begin{figure}
\begin{center}
\leavevmode
\epsfxsize8.6cm
\epsfbox{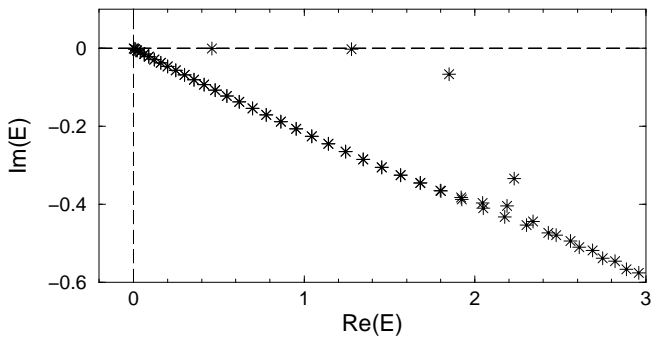}
\end{center}
\caption{
  Spectrum of the complex scaled Hamiltonian (\ref{eq:h0rot}) for $\alpha = 0.1$.
  The stars represent the complex eigenvalues of $H_0^{(\theta)}$ for $\theta = 0.1$,
  which were calculated with a grid basis covering the range $-100 \leq x \leq 100$.
  Four resonances are fully uncovered at this value of the rotation angle.
  \label{fg:sp}
}
\end{figure}

The rotation of the continuum axis uncovers the spectral resonances of the
system, which correspond to the poles of the analytical continuation of the
Green function $G = (E - H_0 + \ii \delta)^{-1}$ to the lower half part of the
complex energy plane.
Those resonances turn into discrete complex eigenvalues $E_n = \mu_n - \ii \Gamma_n / 2$ 
under complex dilation, and are represented by normalizable eigenfunctions
$\psi_n^{(\theta)}(x)$ that can be straightforwardly calculated by diagonalizing
$H_0^{(\theta)}$ in any numerical basis.
This is illustrated in Fig.~\ref{fg:sp} which shows the spectrum of the
rotated Hamiltonian (\ref{eq:h0rot}) for $\theta=0.1$.
Besides a ``continuum'' of levels along $E = \epsilon e^{-2 \ii \theta}$ (which appears as
discretized due to the finite basis), four major resonances can be identified
below the real axis.
Starting with an initial state that is very close to one of the resonances
(i.e., which has, in the complex rotated system, a macroscopic overlap with
the associated wavefunction $\psi_n^{(\theta)}$) will therefore lead to a decay from
the well with a rate that is given by the negative imaginary part $\Gamma_n$ of the
corresponding eigenvalue.

\subsubsection{Nonlinear case}

In the nonlinear case, a major complication arises from the fact that the
interaction-induced contribution $g |\psi(x)|^2$ to the Hamiltonian (\ref{eq:h2})
is nonanalytic in $\psi$.
To avoid this complication, it is tempting to make, in a first approach, the
replacement
\begin{equation}
  |\psi(x)|^2 \mapsto \psi^2(x) \label{eq:repl}
\end{equation}
in Eq.~(\ref{eq:h2}), i.e.\ to consider the analytic nonlinear Schr\"odinger
equation
\begin{equation}
  \widetilde{H}(\psi) \psi(x) = E \psi(x) \label{eq:gpa}
\end{equation}
with
\begin{equation}
  \widetilde{H} = H_0 + g \left[\psi(x)\right]^2 \, . \label{eq:ha}
\end{equation}
Applying the complex scaling transformation (\ref{eq:cs}) to 
Eq.~(\ref{eq:gpa}) yields
\begin{equation}
  \widetilde{H}^{(\theta)}(\psi) \psi^{(\theta)}(x) = E \psi^{(\theta)}(x) \label{eq:gparot}
\end{equation}
with
\begin{equation}
  \widetilde{H}^{(\theta)}(\psi) = H_0^{(\theta)} + g_\theta \left[ \psi^{(\theta)}(x) \right]^2
  \label{eq:harot}
\end{equation}
and $H_0^{(\theta)}$ being defined as in Eq.~(\ref{eq:h0rot}).
Here the effective interaction strength is transformed according to
\begin{equation}
  g \mapsto g_\theta = g e^{- \ii \theta} \label{eq:grot}
\end{equation}
in order to compensate the scaling prefactor $e^{i \theta / 2}$ in Eq.~(\ref{eq:cs}).
This additional scaling reflects the fact that $g$ implicitly contains
information about the norm of the wavefunction [see Eq.~(\ref{eq:gt})].

The above approach yields meaningful results (i.e.\ results that are related
to the actual time-dependent decay process of the condensate) only if the
wavefunction $\psi(x)$ to be calculated in Eq.~(\ref{eq:gp}) is known to be
{\em entirely real}.
In general, this is not the case for unbound resonance states, which typically
exhibit outgoing boundary conditions (i.e, $\psi(x) \propto \exp(\ii k |x|)$ for 
$|x| \to \infty$) and are therefore intrinsically complex.
For such states, the replacement (\ref{eq:repl}) represents a major
modification of the problem, and the attempt to compute the decay behaviour of
the condensate by means of Eqs.~(\ref{eq:gparot}--\ref{eq:grot}) (which, in
effect, were also used in Ref.~\cite{MoiCed05PRA}) would lead to a prediction
that is not expected to be in agreement with alternative approaches based,
e.g., on real-time propagation of the Gross-Pitaevskii equation in presence of
complex absorbing potentials.

To account for the fact that $\psi$ is complex-valued, we formally introduce a
second analytic wavefunction $\overline{\psi}$ which coincides with the complex
conjugate of $\psi$ on the real axis, i.e.
\begin{equation}
  \overline{\psi}(x) \equiv \psi^*(x) \qquad \mbox{for real $x$.} \label{eq:psibar}
\end{equation}
The complex scaling transformation is now defined for $\overline{\psi}$ in the
same way as for $\psi$,
\begin{equation}
  \overline{\psi}(x) \mapsto  \overline{\psi}^{(\theta)}(x) \equiv 
  R_\theta \overline{\psi} (x) = e^{i \theta / 2 } \overline{\psi}( x e^{i \theta} ) \label{eq:csbar}
\end{equation}
with $R_\theta$ the dilation operator.
The analytic continuation of the stationary Gross-Pitaevskii equation
to the complex domain yields then 
\begin{equation}
  H^{(\theta)}(\psi) \psi^{(\theta)}(x) = E \psi^{(\theta)}(x) \label{eq:gprot}
\end{equation}
where the complex scaled Hamiltonian is given by
\begin{equation}
  H^{(\theta)}(\psi) = H_0^{(\theta)} + g_\theta \overline{\psi}^{(\theta)}(x) \psi^{(\theta)}(x)
  \label{eq:hrot} 
\end{equation}
with $g_\theta = g e^{- \ii \theta}$.
Note that $\overline{\psi}^{(\theta)}(x)$ is, in general, {\em not} identical to 
$[{\psi^{(\theta)}}]^*(x)$, the complex conjugate of $\psi^{(\theta)}$. 
It can, however, be obtained from $\psi^{(\theta)}$ via the relation
\begin{equation}
  \overline{\psi}^{(\theta)}(x) = R_\theta \left( \overline{ R_{-\theta} \psi^{(\theta)} } \right) (x)
  \, . \label{eq:psibarpsi}
\end{equation}

For weak or moderate nonlinearities (i.e., with $g$ being of the order of
unity or smaller), the spectrum of self-consistent solutions of the complex
scaled Gross-Pitaevskii Hamiltonian (\ref{eq:hrot}) can be assumend to be not
much different from the linear case shown in Fig.~\ref{fg:sp}:
besides the continuum, a few quasibound resonance levels are expected to arise
below the real energy axis.
In the context of the actual time-dependent decay process of the condensate,
we are mainly interested in the energetically \emph{lowest} resonance, i.e.\
the one with the lowest real part $\mu$ of the complex eigenenergy
$E = \mu - \ii \Gamma/2$ (which would adiabatically evolve into the self-consistent
ground state of the potential well if the barrier height was raised to
infinity).
We shall explain now how the eigenfunction associated with this lowest
resonance can be numerically calculated.

\subsection{Numerical calculation of the resonance state}

\label{s:n}

Our approach is based on the assumption that the lowest resonance state
exhibits a decay rate that is not much larger than the decay rate of the
lowest numerical continuum states.
This is indeed true in the linear case, for the spectrum that is shown in
Fig.~\ref{fg:sp}: 
the width of the lowest resonance is found to be $\Gamma/2 \simeq 10^{-6}$, while the
energetically lowest ``continuum'' state obtained from the numerical
diagonalization (using a grid that covers the spatial range 
$-100 \leq x \leq 100$) decays with the rate $\Gamma/2 \sim 10^{-4}$.
This observation suggests a \emph{real-time propagation} approach to calculate
the lowest resonance state.
We start from a good initial approximation $\psi_0^{(\theta)}(x)$ for the resonant state
(e.g.\ the harmonic eigenstate in the well), and numerically propagate
$\psi^{(\theta)}$ under the \emph{time-dependent} Gross-Pitaevskii equation
\begin{equation}
  \ii \frac{\partial}{\partial \tau} \psi_\tau^{(\theta)}(x) = H^{(\theta)}(\psi_\tau) \psi_\tau^{(\theta)}(x) \label{eq:gptrot}
\end{equation}
in the rotated system
($\tau$ is here a fictitious numerical ``time'' parameter, which is unrelated to
the physical propagation time $t$ in the actual decay process).
In the above linear case, this propagation is guaranteed to converge towards
the eigenstate with the lowest decay rate.

This approach works also for the nonlinear case, but requires some nontrivial
modifications there.
On the one hand, the self-consistent eigenstates and eigenenergies of the
Gross-Pitaevskii equation are defined with respect to a given normalization of
the wavefunction.
A rescaling of $\psi^{(\theta)}$ according to the condition (\ref{eq:norm2}) is
therefore required after each propagation step under Eq.~(\ref{eq:gptrot}).
On the other hand, $\overline{\psi}_\tau^{(\theta)}(x)$ needs to be calculated from
$\psi_\tau^{(\theta)}(x)$ in order to compute the nonlinear term in $H^{(\theta)}(\psi_\tau)$.
In order to implement the prescription (\ref{eq:psibarpsi}) which achieves
this goal, the complex rotation (\ref{eq:cs}) of the wavefunction needs to be
explicitly performed.
This operation, however, is known to be highly unstable \cite{BucGreDel94JPB}
and requires great care in the numerical implementation.

In practice, two different basis sets are simultaneously used in order to
numerically integrate Eq.~(\ref{eq:gptrot}).
To perform the propagation, the rotated wavefunction is expanded in a
\emph{grid basis} --- i.e.,
\begin{equation}
  \psi_t^{(\theta)}(x) = \sum_{n=-n_{\rm max}}^{n_{\rm max}} C_n(t) \chi_n(x) \label{eq:psigrid}
\end{equation}
with 
\begin{equation}
  \chi_n(x) = \left\{ \begin{array}{r@{\; : \;}l}
      1 / \Delta_x & (n - 1/2) \Delta_x < x < (n + 1/2) \Delta_x \\ 0 & \mbox{otherwise}
    \end{array} \right.
\end{equation}
where $\Delta_x$ is a suitable grid spacing.
With the standard finite-difference approximation for the operator of the
kinetic energy,
\begin{equation}
  -\frac{1}{2} \frac{\partial^2}{\partial x^2} \chi_n(x) \equiv 
  - \frac{\chi_{n+1}(x) + \chi_{n-1}(x) - 2 \chi_n(x)}{2 \Delta_x^2} \, ,
\end{equation}
we obtain a tridiagonal Hamiltonian matrix.
This matrix is used to propagate the wavefunction according to the implicit
scheme
\begin{equation}
  \psi_{\tau+ \delta \tau}^{(\theta)} = {\textstyle \left(1 + \ii \frac{\delta \tau}{2} H^{(\theta)}(\psi_\tau) \right)^{-1}}
  {\textstyle \left( 1 - \ii \frac{\delta \tau}{2} H^{(\theta)}(\psi_\tau) \right)} \psi_\tau^{(\theta)} \, .
  \label{eq:prop}
\end{equation}
The latter requires direct matrix-vector multiplications as well as the
solution of linear systems of equations with tridiagonal matrices, which is
efficiently performed by LR decompositions.

For the calculation of $\overline{\psi}^{(\theta)}$, a second, nonorthogonal basis set
is introduced in terms of the analytic \emph{Gaussian} functions
\begin{equation}
  \phi_\nu(x) = \frac{1}{\sqrt{\sqrt{\pi} \sigma_\nu}} \exp\left( -\frac{(x-x_\nu)^2}{2 \sigma_\nu^2}
  \right) \label{eq:gs}
\end{equation}
for $-\nu_{\rm max} \leq \nu \leq \nu_{\rm max}$.
The centers $x_\nu$ and widths $\sigma_\nu$ are chosen according to
\begin{equation}
  x_\nu = \frac{\sinh(\nu \gamma)}{\sinh(\gamma)} \tilde{\Delta}_x \, , \label{eq:gsx}
\end{equation}
for suitable parameters $\gamma,\tilde{\Delta}_x > 0$, and
\begin{equation}
  \sigma_\nu = x_{|\nu+1|} - x_{|\nu|} \, . \label{eq:gss}
\end{equation}
The exponential increase of the widths $\sigma_\nu$ with $|\nu|$, which results from
Eqs.~(\ref{eq:gsx}) and (\ref{eq:gss}), is required to ensure stable
evaluation of linear combinations of $\phi_\nu$ in the complex domain.
$\nu_{\rm max}$ is chosen such that the whole spatial range spanned by the grid
basis is covered by the Gaussians.

From the overlap integrals $\mathcal{V}_{n,\nu} = \int \chi_n(x) \phi_\nu(x) dx$ and 
$\mathcal{I}_{\nu,\nu'} = \int \phi_\nu(x) \phi_{\nu'}(x) dx$, we can determine the expansion
coefficients $D_{- \nu_{\rm max}} \ldots D_{\nu_{\rm max}}$ of $\psi_t^{(\theta)}(x)$ with respect to the
Gaussian basis:
\begin{equation}
  \psi_t^{(\theta)}(x) = \sum_{\nu=-\nu_{\rm max}}^{\nu_{\rm max}} D_\nu(t) \phi_\nu(x) \, . \label{eq:psigauss}
\end{equation}
This involves again the solution of a linear system of equations, namely
$ \sum_{\nu'}\mathcal{I}_{\nu,\nu'} D_{\nu'} = \sum_n \mathcal{V}_{n,\nu} C_n$,
which is accomplished by the LR decomposition of the matrix
$(\mathcal{I}_{\nu,\nu'})$ (the latter is effectively banded, due to the fact that
the Gaussians fall off rapdily with increasing distance from their center).
In a very similar way, we determine the rotation of the wavefunction --- i.e.\
the expansion of $(R_{-\theta}\psi_t^{(\theta)})(x)$ in the (unrotated) Gaussian basis
$(\phi_\nu)$ --- by means of the overlap integrals $\mathcal{I}_{\nu,\nu'}$ and 
$\mathcal{J}_{\nu,\nu'} = \int \phi_\nu(xe^{i\theta}) \phi_{\nu'}(x) dx$.
Complex conjugation of the coefficients, the rotation $R_\theta$, and the
transformation back to the grid basis $(\chi_n)$ finally lead to the state vector
$(\tilde{C}_n)$ that is associated with $\overline{\psi}^{(\theta)}(x)$ \cite{rem_rot}.

We should note that the above method cannot properly reproduce
$\overline{\psi}^{(\theta)}(x)$ far away from the center of the potential.
This is due to the exponential increase of the widths and spacings of the
Gaussians, which prevents a perfect representation of rapidly oscillating
wavefunctions far away from the origin.
We remark, however, that in the time-dependent Gross-Pitaevskii equation 
(\ref{eq:gptrot}) $\overline{\psi}^{(\theta)}(x)$ is multiplied by the square of
$\psi^{(\theta)}(x)$, and the term $(\overline{\psi}^{(\theta)}\psi^{(\theta)}\psi^{(\theta)})(x)$ decreases
exponentially with $|x|$.
Hence, apart from the special case of very large $g$ and rapidly decaying
resonance states, we obtain an approximation for $\overline{\psi}^{(\theta)}$ which is
reasonably good in the relevant spatial regime to ensure stable convergence of
our algorithm.

%% file: csbec_rs.tex
\label{s:r}

\subsection{Chemical potentials and decay rates}

\begin{figure}
\begin{center}
\leavevmode
\epsfxsize8.6cm
\epsfbox{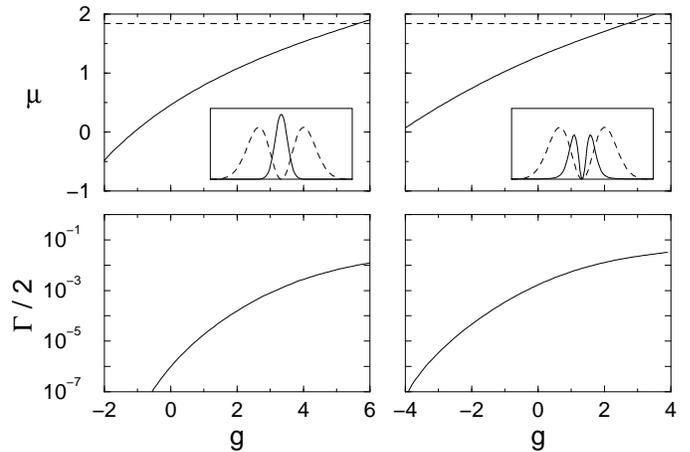}
\end{center}
\caption{Chemical potentials (upper panels) and decay rates (lower panels) of
  the lowest resonance state with even and odd parity (left and right column,
  respectively), as a function of the effective interaction strength $g$.
  The inserts show the spatial density of the corresponding resonance states
  at $g=1$ (with the trapping potential indicated by dashed lines).
  Converged calculations of $\mu$ and $\Gamma$ can be performed up to interaction
  strengths $g$ where the chemical potential reaches the maximum barrier
  height of the potential, marked by the horizontal dashed lines in the upper
  panels.
  For $g < -1.1$, the lowest even state becomes stable, which manifests itself
  in a negative chemical potential and a vanishing decay rate. 
\label{fg:ew}
}
\end{figure}

Fig.~\ref{fg:ew} shows the chemical potentials and decay rates of the
energetically lowest and of the first excited (soliton-like) quasibound state 
as a function of $g$ (lower and upper solid curves, respectively) for $\alpha =
0.1$.
Due to the symmetry of the potential, both states can be calculated with the
implicit propagation scheme described above, where the parity of the state is
selected by the choice of the initial wavefunction.
Typical parameters employed in the calculations are $\Delta_x \simeq 0.01$ for the grid
spacing, $-100 \leq x \leq 100$ for the spatial range covered by the grid, 
$\theta = 0.05$ for the complex scaling angle, and $\gamma \simeq 0.05$ for the parameter
characterizing the exponential increase of the width of the Gaussians.

In the regime of small and moderate nonlinearities ($|g| \lesssim 1$), the numerical
procedure described in the previous subsection is further optimized:
To avoid time-consuming rotations of wavefunctions as best as possible,
the renormalization of $\psi^{(\theta)}$ as well as the calculation of the nonlinear
term in the Gross-Pitaevskii Hamiltonian are performed only every
100th propagation step according to Eq.~(\ref{eq:prop}).
We verified, in any case, the convergence of the calculation by a criterion
that is independent of the real-time propagation approach, namely by the
requirement that the norm of $\delta \psi^{(\theta)} = H^{(\theta)}(\psi) \psi^{(\theta)} - E \psi^{(\theta)}$ fall
below a given precision limit.
To obtain highly accurate results for the chemical potentials and the decay
rates $\Gamma$, it is convenient to perform, for a given value of $g$, several
calculations with different numerical grid spacings $\Delta_x$ (where the total
extent of the grid is kept constant), and to extrapolate $\mu$ and $\Gamma$ in the
limit $\Delta_x \to 0$.
In this way, rather low decay rates down to $\Gamma \simeq 10^{-7}$ can be reproduced.

Since no assumption about the sign of $g$ was made in our approach, we can
also calculate resonance (and bound) states for the case of {\em attractive} 
interaction.
As we see in Fig.~\ref{fg:ew}, the lowest even state of the trapping potential
undergoes, for increasing attraction between the atoms, a transition from a
resonance to a bound state, which is manifested by a negative chemical
potential and a vanishing decay rate.
In agreement with the calculation by Moiseyev et al.\ \cite{MoiO03JPB}, which
were performed for the same potential (\ref{eq:v}) with $\alpha = 0.2$, we find
that this transition occurs at $g \simeq -1.1$.

For strong nonlinearities ($|g| > 1$), the renormalization of $\psi^{(\theta)}$ and
the update of the nonlinear term need to be performed after each propagation
step in order to ensure stable convergence of our approach.
A further modification is introduced in the regime of large decay rates 
$\Gamma \sim 0.01$, i.e.\ where $\mu$ approaches the height of the potential barriers:
To avoid unwanted convergence to energetically low continuum states 
(with decay rates $\Gamma \sim 10^{-4}$, see Sec.~\ref{s:c}), we replace, in this
regime, the real-time propagation approach by a propagation along a
\emph{complex} time path, given by $\tau = \tilde{\tau}e^{i\varphi}$ with real $\tilde{\tau}$
and $0 < \varphi < 2 \theta$.
A stationary solution $\psi$ of the Gross-Pitaevskii equation will then evolve
according to $|\psi_\tau^{(\theta)}(x)| = \exp[( \mu \varphi - \Gamma/2)\tilde{\tau}] \, |\psi_0^{(\theta)}(x)|$ for
small $\varphi$, and quasibound states with finite $\mu$ will thereby be more strongly
enhanced than low-energy continuum states close to the threshold.
Indeed, we find that this modification allows converged calculations of
resonances with chemical potentials close to the barrier height.

\begin{table}
\begin{tabular}{c|ccc|ccc}
  $g$ & $\mu_{\rm CS}$ & $\mu_{\rm ABC}$ & $\mu_{\rm CAP}$ & 
  $\Gamma_{\rm CS}/2$ & $\Gamma_{\rm ABC}/2$& $\Gamma_{\rm CAP}/2$ \\
  \hline
  0 & 0.4601 & 0.4602 & 0.4602 & 9.35e-7 & 9.55e-7 & 9.62e-7 \\
  1 & 0.7954 & 0.7955 & 0.7954 & 1.82e-5 & 1.81e-5 & 1.80e-5 \\
  2 & 1.0765 & 1.0773 & 1.0772 & 1.55e-4 & 1.56e-4 & 1.56e-4 \\
  3 & 1.3190 & 1.3193 & 1.3192 & 8.05e-4 & 8.04e-4 & 8.05e-4 \\
  4 & 1.5315 & 1.5313 & 1.5312 & 2.76e-3 & 2.75e-3 & 2.75e-3 \\
  5 & 1.7236 & 1.7230 & 1.7231 & 6.65e-3 & 6.66e-3 & 6.63e-3 \\
  6 & 1.9043 & 1.9034 & 1.9035 & 1.24e-2 & 1.23e-2 & 1.23e-2
\end{tabular}
\caption{Chemical potentials and decay rates of the lowest quasibound state,
  calculated with the complex scaling approach (CS) and with real-time
  propagation methods using absorbing boundary conditions (ABC) and
  complex absorbing potentials (CAP).
  \label{tb:comp}
} 
\end{table}

Table \ref{tb:comp} shows a comparison of the chemical potentials and decay
rates with the ones that are obtained from real-time propagation approaches
based on the original (i.e., unrotated) Gross-Pitaevskii equation 
(\ref{eq:gp}).
In those calculations, the decay through the boundaries of the numerical grid
is accounted for by absorbing boundary conditions
\cite{Shi91PRB,PauRicSch05PRL} as well as by a complex absorbing potential
of the form $V_{\rm abs}(x) = - \ii ( x / 100 )^{10}$ (for the spatial range
$-100 \leq x \leq 100$).
As in the complex scaling approach, the wavefunction was propagated by an
implicit scheme like (\ref{eq:prop}) (now under the {\em real}, unscaled
Gross-Pitaevskii Hamiltonian) and renormalized after each step to satisfy
the condition (\ref{eq:norm}).
The agreement between the three methods is reasonably good.

We should note, however, that strongly deviating values for $\Gamma$ would
generally be obtained if the resonance states were calculated according
Eqs.~(\ref{eq:gparot},\ref{eq:harot}) under the assumption that the
wavefunction $\psi$ is entirely real.
For $g = 1$, e.g., the real-time propagation approach (\ref{eq:prop}) using
$\widetilde{H}^{(\theta)}$ [from Eq.~(\ref{eq:harot})] instead of $H^{(\theta)}$ yields
essentially the same chemical potential $\mu \simeq 0.795$, but a different decay
rate $\Gamma/2 \simeq 5 \times 10^{-5}$.
Also for $g = 5$ we obtain $\Gamma/2 \simeq 1.6 \times 10^{-2}$ which considerably
overestimates the actual rate $\Gamma/2 \simeq 6.7 \times 10^{-3}$.
The above calculations based on $\widetilde{H}^{(\theta)}$ were performed under the
slightly modified side condition $\int_{-x_\alpha}^{x_\alpha} [\psi(x)]^2 dx = 1$, which would
be consistent with Eq.~(\ref{eq:norm2}) for purely real $\psi$ [the propagation
of $\psi^{(\theta)}$ under $\widetilde{H}^{(\theta)}$ does not seem to converge with the
original side condition (\ref{eq:norm2})].

\subsection{Wavefunctions}

\begin{figure}[t]
\begin{center}
\leavevmode
\epsfxsize8.6cm
\epsfbox{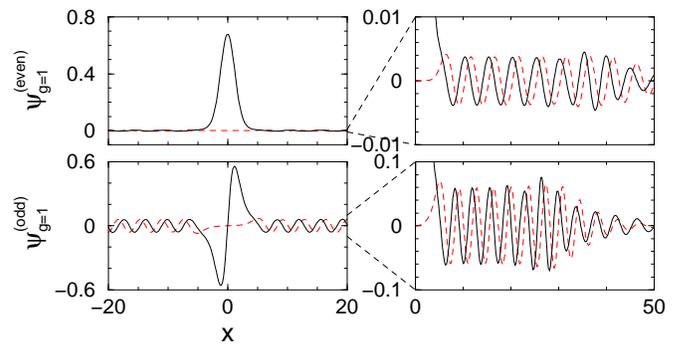}
\end{center}
\caption{(Color online) Wavefunctions of the even and odd resonance states at
  $g=1$ (upper and lower panels, respectively; solid lines: real part; dashed
  lines: imaginary part).
  The wavefunctions were calculated for the parameters $\gamma = 0.05$ and
  $\tilde{\Delta}_x = 0.1$ of the Gaussian basis set (see Eq.~(\ref{eq:gsx})).
  The magnifications in the right panels show that the outgoing tails of the
  wavefunctions are faithfully reproduced till $x \simeq 30$.
  \label{fg:wf1}}
\end{figure}

While the chemical potentials $\mu$ and decay rates $\Gamma$ are apparently well
reproduced by the method of complex scaling, the wavefunctions $\psi(x)$ of the
resonance states (defined according to Eq.~(\ref{eq:gp2}), i.e., in the
unrotated framework) can be calculated only in a limited spatial regime in the
vicinity of the potential well.
We attribute this to the exponential increase of the widths $\sigma_\nu$ and centers
$x_\nu$ with $|\nu|$, which was introduced in order to ensure stable back-rotation
to the real domain.
As already mentioned, this exponential scaling prevents a perfect
representation of rapidly oscillating wavefunctions far away from the origin.
For low and moderate nonlinearities ($g \lesssim 1$), we typically reproduce $\psi(x)$
up to distances of the order of $|x| \simeq 50$ from the center of the potential.
This is illustrated in Fig.~\ref{fg:wf1} where the wavefunctions of the even
and odd resonance states at $g=1$ are plotted.

\begin{figure}[t]
\begin{center}
\leavevmode
\epsfxsize8.6cm
\epsfbox{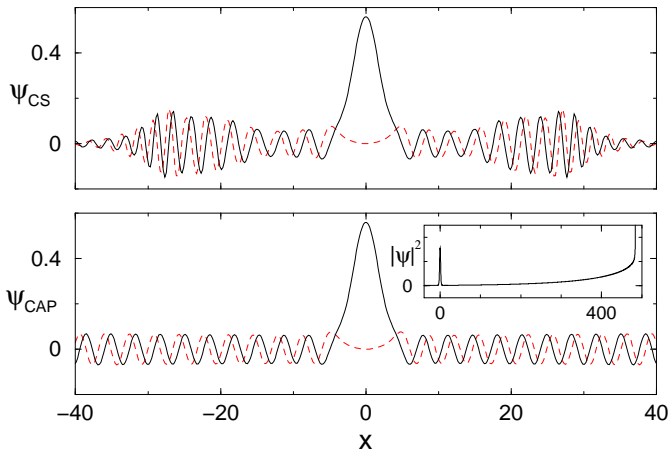}
\end{center}
\caption{(Color online) Wavefunction of the even resonance state at $g=5$
  (solid line: real part; dashed line: imaginary part), calculated with the
  complex scaling method (upper panel) and with a real-time propagation
  approach using complex absorbing potentials (lower panel).
  The comparison shows that the complex scaling approach fails to correctly
  reproduce the resonance wavefunction for $|x| > 15$, even though
  the chemical potential and the decay rate are in agreement for both methods 
  (see Table \ref{tb:comp}).
  The insert in the lower panel displays the density of the resonance state
  continued till $x = 500$, the tail of which was computed by integrating the
  free one-dimensional Gross-Pitaevskii equation (with $v(x) \equiv 0$) with the
  complex effective chemical potential $\mu_{g=5} - \ii \Gamma_{g=5} / 2$, starting from
  $\psi_{\rm CAP}(x=10)$.
  A singularity of $|\psi|^2$ is encountered at $x \simeq 485$ where the density
  exceeds the critical value $\mu / g$.
  \label{fg:wfc}}
\end{figure}

For large $g$, the spatial range in which $\psi$ can be correctly reproduced
becomes shorter.
This is demonstrated in the upper panel of Fig.~\ref{fg:wfc} which displays
the wavefunction $\psi_{\rm CS}$ of the even resonance state at $g = 5$
calculated with the complex scaling approach.
In contrast to the actual resonance wavefunction (shown in the lower panel of
Fig.~\ref{fg:wfc}), the density $|\psi_{\rm CS}(x)|^2$ exhibits unphysical maxima
near $|x| \simeq 25$ and decreases rapidly for larger $|x|$.
We believe that those features arise due to the imperfect computation of the
complex conjugate function $\overline{\psi}^{(\theta)}(x)$ by means of explicit
rotations within the Gaussian basis set.
It should be noted, however, that the values for $\mu$ and $\Gamma/2$ that are
extracted from this wavefunction are nevertheless in good agreement with the
chemical potentials and decay rates calculated with alternative methods at 
$g = 5$ (see Table \ref{tb:comp}).

Let us finally remark that from a formal point of view, even the ``true''
resonance wavefunction cannot be defined within an infinitely large spatial
regime for $g > 0$:
Due to the outgoing boundary conditions, the density $|\psi(x)|^2$ of the
resonance wavefunction constantly increases with increasing $|x|$ outside the
potential well, until it exceeds the critical value $\mu / g$ where the
kinteic energy formally vanishes.
Beyond that point, a rather sudden increase of the density towards infinity is
encountered \cite{rem_div}, which indicates that permanently time-dependent
behaviour would be expected for the actual decay process in that spatial
regime.
We should note, however, that the associated critical distance $|x| = x_c$ at
which $|\psi(x)|^2$ equals $\mu / g$ is much larger than the spatial regime taken
into account in our numerical calculations; 
for the even resonance state at $g = 5$, e.g., we obtain $x_c \sim 500$ from 
$\mu \simeq 1.7$ and $\Gamma/2 \simeq 0.007$, while even larger critical distances 
($x_c \sim 10^6$) would be expected for the more long-lived resonance states
around $g = 1$.

\subsection{The time-dependent decay process}

\begin{figure}[t]
\begin{center}
\leavevmode
\epsfxsize8.6cm
\epsfbox{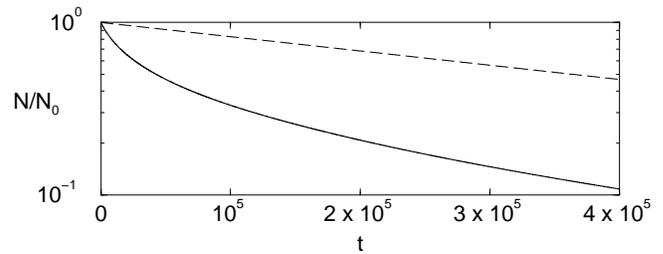}
\end{center}
\caption{Time-dependent decay of the condensate. Plotted is the number $N$ of
  atoms inside the potential well as a function of time, with $N_0 \equiv N(t=0)$
  such that $g = 1$ initially.
  The dashed line shows the corresponding decay behaviour of noninteracting
  atoms.
  \label{fg:dec}}
\end{figure}

With the above method to calculate the decay rates of the resonance states for
given nonlinearity strengths $g$, we are now in a position to predict the 
actual, real-time decay process of the condensate.
As was derived in Section \ref{s:p} from the adiabatic approximation
(\ref{eq:psi_ad}), the number of atoms in the potential well decreases with
time according to
\begin{equation}
  \frac{d}{dt} N(t) = - \Gamma_{g(t)} \, N(t) \, , \label{eq:dN2}
\end{equation}
where $g(t)$ is proportional to $N(t)$ via Eq.~(\ref{eq:gt}):
\begin{equation}
  g(t) = 2 \frac{a_s}{a_0}\frac{\omega_\perp}{\omega_0} N(t) \, . \label{eq:g2}
\end{equation}
Multiplying Eq.~(\ref{eq:dN2}) with the constant prefactors appearing in
Eq.~(\ref{eq:g2}) yields the ordinary differential equation
\begin{equation}
  \frac{d}{dt} g(t) = - \Gamma_{g(t)} \, g(t) \, . \label{eq:dg}
\end{equation}
This equation can be efficiently integrated e.g.\ with a Runge-Kutta solver,
provided the rates $\Gamma_g$ are known in the range $0 \leq g \leq g(0)$.

Fig.~\ref{fg:dec} shows the result of this integration for the case of
repulsive interactions ($a_s > 0$).
Here, the initial number of atoms $N(0)$ was assumed such that $g(0) = 1$.
The calculation was based on the decay rates $\Gamma_0$, $\Gamma_{0.1}$ \ldots $\Gamma_1$ that
were computed with the complex scaling method, and employed cubic
interpolation to obtain intermediate values for $\Gamma_g$.
We clearly see that $N(t)$ decays initially according to a subexponential law,
due to the fact that $\Gamma$ decreases with decreasing $g$
(a superexponential law would be encountered for attractive interaction).
After about $10^5$ time units, the exponential decay behaviour of a
noninteracting condensate (with decay rate $\Gamma_0 \simeq 2 \times 10^{-6}$) is recovered.

For comparison, we also performed a full time-dependent calculation of the
decay process of the condensate.
To this end, we first computed, using again Eq.~(\ref{eq:norm2}) as side
condition, the self-consistent ground state of the condensate at $g = 1$ in
the ``closed'' trapping potential
\begin{equation}
  \bar{v}(x) = \left\{ \begin{array}{r@{\; : \;}l} v(x) & |x| < x_\alpha \\
      v(x_\alpha) & |x| > x_\alpha \end{array} \right.
\end{equation}
with $x_\alpha = 1 / \sqrt{\alpha}$, where the exterior part of $v(x)$ is replaced by a
constant level corresponding to the height of the barriers.
At $t = 0$, $\bar{v}(x)$ was suddenly ``opened'', i.e.\ replaced by $v(x)$,
and the wavefunction of the condensate's ground state was propagated under the
time-dependent Gross-Pitaevskii equation at $g=1$, where absorbing boundary
conditions were used to account for the decay through the boundaries of the
numerical grid.
The agreement between the two approaches is extremely good.
At $t = 4 \times 10^5$ for instance, when the surviving population of the
condensate has decayed to $N(t) \simeq 0.1 N_0$, we find that the relative
difference between the above propagation method and the integration of
Eq.~(\ref{eq:dg}) (which takes much less CPU time, including the calculation
of the decay rates $\Gamma_g$) is of the order of $\Delta N/N(t) \simeq 0.01$.
This shows that the rates $\Gamma_g$ obtained from the complex scaling approach are
sufficiently precise to predict the time-dependent decay dynamics of the
condensate.

%% file: csbec_cc.tex
In summary, we have shown that the method of complex scaling can be used to
calculate metastable mean-field states of Bose-Einstein condensates that are
confined in trapping potentials with finite tunneling barriers.
Our approach is based on the complex rotation $x \mapsto x e^{\ii \theta}$ of the
Gross-Pitaevskii equation and employs, as key ingredients, separate dilations
of the condensate wavefunction $\psi$ and its conjugate $\overline{\psi}$, together
with an additional scaling $g \mapsto g e^{- i \theta}$ of the interaction strength.
A real-time propagation approach is used to calculate the lowest symmetric
and antisymmetric quasibound states of the trapping potential and to determine
the associated chemical potentials and decay rates.
We find good agreement with alternative propagation methods that are based on
the unrotated Gross-Pitaevskii equation and use absorbing boundary conditions
as well as complex absorbing potentials to describe the decay.

Though only exemplified for the harmonic trap with Gaussian cutoffs, our
method is sufficiently general to be applied to other trapping geometries,
such as the double barrier potentials that are proposed in the context of
resonant transport \cite{PauRicSch05PRL,CarLar99PRL}.
This is indeed confirmed in a recent study on the nonlinear Wannier-Stark
problem, where the decay rates of the condensates were reproduced with the
complex scaling method \cite{WimSchMan05JPBs}.
The formalism presented here can, in addition, be straightforwardly
generalized to describe decay processes in two- and three-dimensional
geometries (which would naturally involve a higher numerical effort).
Moreover, the framework of complex scaling seems also suited to be applied
beyond the pure mean-field description of the condensate, e.g.\ in the context
of fragmentation \cite{CedStr03PLA} and within the microscopic quantum
dynamics approach \cite{KoeBur02PRA}.
We therefore expect that the complex scaling technique might develop into a
useful tool to predict the storage time of condensates in microscopic trapping
potentials, and to address the issue of resonances of the nonlinear
Gross-Pitaevskii equation from the conceptual point of view.

%% file: csbec.bbl
\begin{thebibliography}{10}
\bibitem{rem_rot}
The rotation of $\psi^{(\theta)}$ back to the real domain is also employed for
  the determination of the factor by which the wavefunction has to be rescaled
  after each propagation step, in order to satisfy the normalization condition
  (\ref{eq:norm2}).
\bibitem{rem_div}
This behaviour can be straightforwardly reproduced by integrating, with a
  Runge-Kutta solver, the ordinary differential equation corresponding to the
  stationary Gross-Pitaevskii equation (\ref{eq:gp2}) in presence of a complex
  energy $E = \mu - \ii/2 \Gamma$.
\end{thebibliography}
